\documentstyle[12pt,aaspp4,flushrt]{article}
\input{psfig}

\makeatletter

\@ifundefined{chapter}{\def\thebibliography#1{\section*{References\@mkboth
  {REFERENCES}{REFERENCES}}\list
  {\relax}{\setlength{\labelsep}{0em}
        \setlength{\itemindent}{-\bibhang}
        \setlength{\itemsep}{0pt}
        \setlength{\parsep}{0pt}
        \setlength{\leftmargin}{\bibhang}}
    \def\newblock{\hskip .11em plus .33em minus .07em}
    \sloppy\clubpenalty4000\widowpenalty4000
    \sfcode`\.=1000\relax}}%
{\def\thebibliography#1{\chapter*{Bibliography\@mkboth
  {BIBLIOGRAPHY}{BIBLIOGRAPHY}}\list
  {\relax}{\setlength{\labelsep}{0em}
        \setlength{\itemindent}{-\bibhang}
        \setlength{\itemsep}{0pt}
        \setlength{\parsep}{0pt}
        \setlength{\leftmargin}{\bibhang}}
    \def\newblock{\hskip .11em plus .33em minus .07em}
    \sloppy\clubpenalty4000\widowpenalty4000
    \sfcode`\.=1000\relax}}

\newlength{\bibhang}
\let\@internalcite\cite
\def\cite{\let\@citeleft(\let\@citeright)%
    \@ifstar{\citeyear}{\citefull}}
\def\citenp{\let\@citeleft\relax\let\@citeright\relax
    \@ifstar{\citeyear}{\citefull}}
\def\citefull{\def\astroncite##1##2{##1~##2}\@internalcite}
\def\citeyear{\def\astroncite##1##2{##2}\@internalcite}

\def\@citex[#1]#2{\if@filesw\immediate\write\@auxout{\string\citation{#2}}\fi
  \def\@citea{}\@cite{\@for\@citeb:=#2\do
    {\@citea\def\@citea{; }\@ifundefined
       {b@\@citeb}{{\bf ?}\@warning
       {Citation `\@citeb' on page \thepage \space undefined}}%
{\csname b@\@citeb\endcsname}}}{#1}}

\def\@cite#1#2{\@citeleft#1\if@tempswa , #2\fi\@citeright}
\def\@biblabel#1{}

\makeatother

\newcommand\approxgt{\mbox{$^{>}\hspace{-0.24cm}_{\sim}$}}
\newcommand\approxlt{\mbox{$^{<}\hspace{-0.24cm}_{\sim}$}}

\begin{document}

\title{A Geometrical Test of the Cosmological Energy Contents Using the
Lyman-alpha Forest} 
\author{Lam Hui\altaffilmark{1}, Albert Stebbins\altaffilmark{1} and
Scott Burles\altaffilmark{2}}
\altaffiltext{1}{NASA/Fermilab Astrophysics Center, Fermi
National Accelerator Laboratory, Batavia, IL 60510; e-mail: \it
lhui@fnal.gov, \it stebbins@fnal.gov}
\altaffiltext{2}{Department of Astronomy and Astrophysics, The University
of Chicago, Chicago, IL 60637; e-mail: \it 
scott@oddjob.uchicago.edu}

\begin{abstract}
In this {\it Letter} we explore a version of the test of
cosmological geometry proposed by Alcock \& Paczy\'nski \cite*{ap79}, using
observations of the Lyman-$\alpha$ forest in the spectra of close quasar
pairs.  By comparing the correlations in absorption in one quasar spectrum
with correlations between the spectra of neighboring quasars one can determine
the relation of the redshift distance scale to the angle distance scale at the
redshift of the absorbers, $z \sim 2 - 4$.  Since this relationship depends on
the parameters of the cosmological model, these parameters may be determined
using the Lyman-$\alpha$ forest.  While this test is relatively insensitive to 
the density parameter $\Omega_m$ in a dust-dominated universe, it is
more sensitive to the presence of a matter component with large
negative pressure 
(such as a cosmological constant $\Lambda$) and its equation of state.
With only $25$ pairs of quasar 
spectra at angular separations $0.5' - 2'$, one can discriminate between an
$\Omega_m = 0.3$ open universe ($\Lambda=0$) and an $\Omega_m = 0.3$ flat
($\Lambda$-dominated) universe at the $4-\sigma$ level. The S/N can be enhanced
by considering quasar pairs at smaller angular separations, but requires proper
modeling of nonlinear redshift space distortions. Here the correlations and
redshift space distortions are modeled using linear theory.
\end{abstract}


\section{Introduction}
\label{intro}

Recent supernova Ia observations have generated a lot of interest in
cosmological models where a significant fraction of the energy
contents has negative pressure
(\citenp{perlmutter97,riess98,garnavich98}). A common way of
parameterizing the equation of state of this component, which we call
$Q$, is $p = w \rho$, where $p$ and $\rho$ denote the pressure
and density respectively.
The cosmological constant $\Lambda$ corresponds to the case $w = -1$.

It is important to have independent ways to constrain the abundance and
properties of such a component, as different methods suffer from different
systematic errors, and, perhaps more importantly, different methods are
sensitive to different combinations of parameters.  In this paper, we discuss a
version of a test proposed by Alcock \& Paczy\'nski (\citenp*{ap79}; AP
hereafter), which is particularly sensitive to the presence/absence of $Q$.
They observed that an object placed at a cosmological distance would have a
definite relationship between its angular and redshift extents, which is
cosmology-dependent.  

Consider an object with mean redshift $\bar z$, and angular size $\theta$ which
thus has transverse extent (in velocity units)
\begin{equation}
u_\perp(\theta) = {\bar H \over {1+\bar z}} D_A (\bar z) \theta \ .
\label{uperp}
\end{equation}
Here $\bar H$ is the Hubble parameter at redshift $\bar z$, and
$D_A (\bar z)$ is the angular diameter distance. 
For spherical objects the radial and transverse extents are equal, but more
generally if the object is squashed radially by a factor $\alpha_s$, then the
radial extent is $u_\parallel \equiv{c\Delta z\over 1+z}=\alpha_s u_\perp$
Here $c$ is the speed of light and $u_\perp,\,u_\parallel\ll c$ is assumed. 
A plot of $u_\perp/\theta$ is shown in Fig. \ref{angQK}. 
Note how its value for a $Q$-dominated universe (for $w\,\approxlt\,-1/3$)
differs significantly from that for a no-$Q$-universe.

Various incarnations of this test have been discussed in the context of galaxy
and quasar surveys (e.g. 
\citenp{ryden95,bph96,matsubara96,popowski98,delaix98}) where the ``object''
used is the two-point correlation function,  whose ``shape''($\alpha_s$
above)  need not be spherical because of redshift-anisotropy induced by
peculiar motion.  In the case of the Lyman-alpha (Ly$\alpha$) forest, we cannot
observe the full 
three-dimensional (3D) correlation directly. Instead we can measure
the one-dimensional (1D) correlation along a line of sight (LOS), and
the cross-correlation 
between two (or multiple) close-by LOS, or their Fourier
counterparts: the auto- and the cross-spectra.
The two are related to each other through an underlying
3D power spectrum. These relations are spelled out in \S
\ref{general} (for auto-spectrum, see also
\citenp{kp91}). Using a method developed by Hui \cite*{hui98} building
on earlier work by Croft et al. \cite*{croft98}, one can invert the
auto-spectrum to obtain the underlying 3D power spectrum, which then
allows one to predict what the cross-spectrum should be, except that
the prediction is cosmology-dependent. A comparison between the
observed and the predicted cross-spectrum is then our version of the
AP test. It is not the only version possible in the context of the
forest, but we will use this version to gain some intuition
on the sensitivity of the AP test to various cosmological parameters.
A hypothetical implementation of this test is studied in \S \ref{AP}.

An important problem in this application of the AP test is the
modeling of the redshift-space distortions (the ``shape''). We give a
first estimate using linear theory in \S 
\ref{perturbAC}. We conclude in \S \ref{conclude} with an
assessment of the expected S/N of the AP-effect-measurement in
the Ly$\alpha$ 
forest, and remarks on lines of further investigation.

\section{Auto- and Cross-spectrum: Definitions and General Formula}
\label{general}

Given the observed transmission $f = e^{-\tau}$ as a function of
the velocity $u_\parallel$ along a LOS and the angular position
$\theta$ on the sky, let us define $\delta_f = (f - \bar f) / \bar f$ where 
$\bar f$ is the mean transmission. Then the auto-correlation
$\xi^f_\parallel$ along the LOS and the 
cross-correlation $\xi^f_{\times}$ between two close-by LOS
of angular separation $\theta$, and their Fourier counterparts the
auto-spectrum $P^f_\parallel$ and the cross-spectrum $P^f_\times$, are
respectively defined by  
\begin{eqnarray}
\label{XiaXic}
&& \langle \delta_f (u_\parallel', \theta') \delta_f
(u_\parallel'+u_\parallel, \theta') \rangle =
\xi^f_\parallel (u_\parallel) \, \, , \, \,  
\langle \delta_f (u_\parallel', \theta') \delta_f (u_\parallel'+u_\parallel,
\theta'+\theta) \rangle = \xi^f_\times (u_\parallel, \theta) \\ 
\label{PaPc}
&& P^f_\parallel (k_\parallel) = \int \xi^f_\parallel (u_\parallel)
e^{-ik_\parallel u_\parallel} du_\parallel \, \, , \, \, 
P^f_\times (k_\parallel, \theta) = \int \xi^f_\times (u_\parallel,
\theta) e^{-ik_\parallel u_\parallel} du_\parallel
\end{eqnarray}
where we use $k_\parallel$ to denote the wave-vector along the line of
sight. 

The two quantities are related to an underlying
3D power spectrum $\tilde P^f (k_\parallel, k)$, where
we use $\tilde{}$ to denote the 3D nature, and $k$ is the
length of the 3D wave-vector:
\begin{eqnarray}
\label{PthreeD}
P^f_\parallel (k_\parallel) &=& \int_{k_\parallel}^\infty \tilde P^f
(k_\parallel, k) k {dk \over 2 \pi} \\ \nonumber
P^f_\times (k_\parallel, \theta) &=& \int_{k_\parallel}^\infty \tilde P^f
(k_\parallel, k) J_0 [k_\perp u_\perp(\theta)]  k {dk \over 
2 \pi} \\ \nonumber  
\end{eqnarray}
where $J_0 [x] = \int_0^{2\pi} e^{-i x \, {\rm cos}\, \alpha}
d\alpha/(2\pi)$ is the zeroth order Bessel function, $k_\perp = \sqrt{k^2 -
k_\parallel^2}$, and $u_\perp$ is the transverse
velocity-separation for the given $\theta$ (eq. [\ref{uperp}]). 

We deliberately allow $\tilde P^f (k_\parallel, k)$ two arguments to
account for 
the possibility of an anisotropic 3D
power spectrum, 
as for instance in the presence of distortions by peculiar motion, or
thermal broadening (bear in mind that line-broadening always acts
along, not transverse, to the LOS). 
The tricky part is how to model this distortion. A general form is
\begin{equation}
\tilde P^f (k_\parallel, k) = W(k_\parallel/k, k) \tilde P^f (k)
\label{W}
\end{equation}
where $W$ is a distortion kernel,
and $\tilde P^f (k)$ with only one 
argument represents the transmission power spectrum in the absence of peculiar
motion and line-broadening.

As we will discuss in \S \ref{perturbAC}, eq. (\ref{W}) together
with eq. (\ref{PthreeD}) allow us to predict the cross-spectrum, once
the auto-spectrum is given and $W$ as well as the relevant
cosmological parameters 
determining the angle-velocity relation (eq. [\ref{uperp}]) are
specified. 
A comparison of the predicted and observed cross-spectra then allows
one to discriminate between different cosmological models, which is
our version of the AP test.
%
We will perform a first estimate of $W$
using perturbation theory.

\section{A Perturbative Estimate}
\label{perturb}

\subsection{Perturbative Auto- and Cross-spectra}
\label{perturbAC}

A linear perturbative calculation (i.e. assume $\delta_f \ll 1$) of
the auto-spectrum has been carried out 
in Hui \cite*{hui98}. The cross-spectrum can be computed in a very
similar manner. We simply state the results here, and refer the reader
to Hui \cite*{hui98} for details.

Essentially, both power spectra follow from the substitution of eq.
(\ref{W}) into eq. (\ref{PaPc}), with $W$ given by\footnote{We have
equated $\tilde P^f (k)$ with the $A' \tilde P^\rho (k)\, {\rm
exp}[-k^2/k_F^2]$ in Hui 
\cite*{hui98}, where $A'$ is a constant, $k_F$ is roughly the Jean
scale, and $\tilde P^\rho (k)$ is
the 3D real-space mass power spectrum.} 
\begin{equation}
W (k_\parallel/k,k) = \Biggl[ 1 + \beta_f {k_\parallel^2 \over
k^2} + \Delta_b \Biggr]^2 \, {\rm exp} [-
{k_\parallel^2/{k_\parallel^s}^2}]
\label{Wperturb}
\end{equation}
with 
\begin{equation}
\beta_f = {f_\Omega \over {2-0.7 (\gamma-1)}} \, \, , \, \, \Delta_b =
{1-\gamma \over {8 - 2.8 (\gamma -1)}} k^2_\parallel b_{T_0}^2
\label{betaDelta}
\end{equation}
where $f_{\Omega} = d\, {\rm ln D}/d\, {\rm ln a}$ with
$D$ being the linear growth factor and $a$ the expansion scale factor, 
$b_{T_0}$ is the thermal broadening width associated with a mean
temperature of $T_0$, and
$\gamma$ specifies the temperature-density relation through $T \propto
\rho^{\gamma-1}$ where $T$ is gas pressure and 
$\rho$ is the density. The smoothing scale $k_\parallel^s$ is associated
with the thermal broadening scale as well as observation resolution.
(see Hui \citenp*{hui98} for details).

An important feature of eq. (\ref{Wperturb}) is that on large scales
($k_\parallel,\,k\ll k_\parallel^s,\,1/b_{T_0}$), it reduces to the famous
Kaiser (\citenp*{kaiser87}) formula, with $\beta_f$ playing the role of the
galaxy-bias factor:
\begin{equation}
W (k_\parallel/k,k) \rightarrow W_{\rm LS} (k_\parallel/k,k) =
(1+\beta_f {k_\parallel^2 /k^2})^2 
\label{Wlinear}
\end{equation}

As shown in Hui \cite*{hui98}, this allows a one-parameter-only
($\beta_f$) inversion of the large scale
3D isotropic power spectrum $\tilde P^f (k)$ using this linear integral
equation, which follows from eq. (\ref{PthreeD}) and (\ref{W}):
\begin{equation}
P^f_\parallel (k_\parallel) = \int_{k_\parallel}^\infty W (k_\parallel/k,k)
{\tilde P^f} (k) {k dk \over {2 \pi}} \, .
\label{projection}
\end{equation}
The above, in discretized form, can be viewed as a matrix equation,
and one simply inverts a matrix proportional to $W$ 
(which is in essence upper or lower triangular because of the limits of
integration/summation) to obtain $\tilde
P^f (k)$ from the auto-spectrum. It can be shown that 
$W_{\rm LS}$ (eq. [\ref{Wlinear}]) could be used
instead of the full $W$ in obtaining the large scale $P^f (k)$, with
good accuracy.
The reader is referred to Hui \cite*{hui98} for illustrations of
this method.

An interesting bonus of the distortion kernel given in eq.
(\ref{Wperturb}) is that the factor of ${\rm exp} [-{k_\parallel^2/
{k_\parallel^s}^2}]$ is commonly used to describe nonlinear redshift
distortions of galaxy distributions (see e.g. \citenp{pd94}).
This should come as no surprise because nonlinear distortions arise
from virialized objects which have a Maxwellian distribution of
velocities -- precisely the form for thermal broadening as
well, which determines $k_\parallel^s$. One can then use the
perturbative kernel in eq. (\ref{Wperturb}) but allow $k_\parallel^s$
to vary to account for nonlinear distortions. We will not pursue that
here, but will come back to it in \S \ref{conclude}. It should be
emphasized that even if $W$ turns out to be substantially different
from eq. (\ref{Wperturb}) or (\ref{Wlinear}), perhaps due to nonlinear
clustering, the procedure of inverting
a triangular matrix using eq. (\ref{projection}) should generally work.

Given the 3D isotropic power spectrum $\tilde P^f (k)$, one
can predict the cross-spectrum using:
\begin{equation}
P^f_\times (k_\parallel,\theta) = \int_{k_\parallel}^\infty W
(k_\parallel/k,k) 
{\tilde P^f} (k) J_0 [k_\perp u_\perp(\theta)]
{k dk \over {2 \pi}} 
\label{projectionC}
\end{equation} 
which follows from eq. (\ref{PthreeD}) and (\ref{W}).
As in the inversion procedure above, we will use $W_{\rm LS}$ instead
of the full $W$, when performing the above integration.
We will see in the next section that the induced error in the 
predicted large scale cross-spectrum is small. Note that
the rapid oscillation of $J_0$ and decay of $\tilde P^f$  at high
$k_\perp$ or $k$ means 
the large scale $P^f_\times$ is not sensitive to assumptions made about
the small scale distortion kernel.

\subsection{The Alcock-Paczy\'nski Test}
\label{AP}

The AP test for the Ly$\alpha$
forest could be implemented in several different ways. 
The version adopted here should be seen as a first step towards
understanding the sensitivity of the test. We will discuss the issue
of other possible implementations in \S \ref{conclude}.

Suppose one is given a set of idealized observed transmission power
spectra, say the auto-spectrum and the cross-spectra for two different
angular separations $\theta = 1'$ and $2'$. 
In real life, one of course does not know {\it a priori}
the underlying cosmological model behind this set of observations.
For the sake of our testing here, let us construct these power spectra
by assuming an input ``true'' model: the open Cold-Dark-Matter (OCDM)
universe with $\Omega_m = 0.3$ and no $Q$.
The full distortion kernel (eq. [\ref{Wperturb}]) is used to compute
these power spectra, with $\gamma = 1.5$, and $b_{T_0}$, $k_\parallel^s$ and
$k_F$ corresponding to a gas of temperature $10^4 \, {\rm K}$
(
see Hui
\citenp*{hui98} for details). The mass power spectrum is assumed
to have a shape parameter of $\Gamma = 0.25$. Let $\bar z = 3$.

Pretending we have no insider
information on the underlying cosmology, our version of the AP test comes
in two steps. First, perform the inversion from the ``observed'' auto-spectrum
$P^f_\parallel (k_\parallel)$ to the 3D isotropic power spectrum
$\tilde P^f (k)$ using eq. (\ref{projection}) and $W_{\rm LS}$ in eq.
(\ref{Wlinear}). By using the 
latter instead of the full kernel as in eq. (\ref{Wperturb}), we will
be making an error. However, the error will be small on sufficiently
large scales, as we will see.
The second step involves predicting the cross-spectra for the
corresponding angular separations, using eq. (\ref{projectionC}), once
the 3D isotropic power spectrum 
$\tilde P^f (k)$ is computed.

Both steps require the assumption of a cosmological model.
In the first step, the cosmological density parameters determine
the amount of redshift-space anisotropy that needs to be accounted
for, through the parameter 
$\beta_f$ or $f_\Omega$ (eq. [\ref{Wlinear}] or
[\ref{betaDelta}]).
The temperature-density-relation index $\gamma$ also enters into
$\beta_f$; we will come back to this below.
In the second step, cosmology dictates the angle-velocity relation
through the quantity $u_\perp (\theta)$ (eq. [\ref{uperp}] \&
[\ref{projectionC}]). 
 
To parameterize the cosmological model that we need to assume in
carrying out the AP test, we use $\Omega_m$, $\Omega_k$ and $\Omega_Q$,
which correspond to the mass, curvature and $Q$ density parameters.
They sum to 1. The $Q$ component is
described by an equation of state of the form $p = w \rho$
where $ -1 \leq w \leq 0$. The limiting cases of $-1$ and $0$
correspond to the cosmological constant $\Lambda$ and matter $m$
respectively. For $w > -1$, the $Q$ component could
cluster (e.g. \citenp{frieman95,ferreira98,caldwell97}), which would
affect the distortion 
parameter $\beta_f$.  
We consider the simple case of no $Q$-clustering here (see e.g.
\citenp{tw97}). 

Assuming any set of parameters different from that of the underlying
input model (OCDM) will result in predicted cross-spectra different
from the ``observed'' ones. The fractional errors, for
a set of 5 cosmological models, are shown in Fig. \ref{pinvC7}.

In the case where the actual input
cosmology is used, the cross-spectra are quite accurately predicted.
It is not perfect, especially at smaller scales, because the
inversion procedure in the first step of our version of the AP test is
inherently approximate by avoiding the modeling of small scale distortions
(see \citenp{hui98}).
However, in cases where the ``wrong'' model-cosmology is assumed, the
predicted cross-spectra are systematically different from the
``observed'' or input cross-spectra.

The effect is most pronounced when $Q$ is present. 
The canonical example is the cosmological constant with $w = -1$,
which gives the 
strongest departure, among the five models, of the predicted
cross-spectra from those of the input cosmology, with 
differences as large as
$200 \%$ at sufficiently large $k_\parallel$'s.
Increasing $w$ while keeping $\Omega_Q$ fixed (the $w = -1/3$ model),
or decreasing $\Omega_Q$ while keeping $w$ fixed (adding in mixture of
curvature, as in the somewhat perverse model of $\Omega_m = 0.2$,
$\Omega_k = 0.4$ \& $\Omega_\Lambda = 0.4$) tends to bring the
predictions closer to that of the input open universe, but still with
substantial differences. The $\Omega_m = 1$ critical matter density
universe is closest to the input model (that is, aside from the input
model itself), with a difference of about $10 - 20 \%$. 

As we discuss before, the cosmological model influences
the prediction of the cross-spectra through two parameters:
$f_\Omega$ (or $\beta_f$; see eq.
[\ref{betaDelta}]) of the distortion 
kernel, and $u_\perp (\theta)$ (eq. [\ref{uperp}]) of the
velocity-distance relation. 
To understand approximately the size of the effect we are measuring,
it is simplest to ignore redshift-distortion first, and consider the
influence of the second parameter alone. 

Putting $W = 1$ and assuming a power-law $\tilde P^f (k) = B
k^{n}$ in eq. (\ref{projectionC}), it can be shown that (\citenp{as64})
\begin{equation}
P^f_\times (k_\parallel,\theta) = {B\over 2 \pi} \left[{2 k_\parallel
/ u_\perp}\right]^{1 + {n\over 2}} \, K_{1+{n\over 2}} (k_\parallel
u_\perp) \, / \, \Gamma (-n/2)
\label{pcrossanalytic}
\end{equation}
where $K_\nu$ is the $\nu$-th order modified Bessel function and
$\Gamma$ is the gamma function. $K_\nu (x)$ has an asymptote
of $\sqrt{\pi/(2x)}\,{\rm exp}[-x]$ in the large-$x$ limit. (
For $\nu =\pm1/2$ this is exact.) For the scales of interest, $n$ should vary
somewhere between $-2$ and $-3$. 
For sufficiently large $k_\parallel$ (i.e.
$k_\parallel u_\perp \, \approxgt \, \pi$), the $u_\perp$ and $k_\parallel$
dependence of 
the cross-spectrum is given by: for $n = -2$, 
$P^f_\times (k_\parallel,\theta) 
\propto \, {\rm exp} 
[-k_\parallel u_\perp] / \sqrt{k_\parallel u_\perp}$, while for $n =
-3$, $P^f_\times (k_\parallel, \theta) \propto \, {\rm exp}
[-k_\parallel u_\perp] / k_\parallel$. 

This exponential dependence on $u_\perp (\theta)$ explains the large
differences between different models at the smaller scales.
(Of course, the exponential suppression of power at high
$k_\parallel$'s also means the cross-spectrum is going to be hard to
measure at such scales. See \S \ref{conclude}.)
We show in Fig. \ref{angQK} two curves depicting how $u_\perp /
\theta$ varies with $w$ and with $\Omega_k$. It can be seen that pure
matter-plus-curvature models (with no $Q$) generally have a higher
$u_\perp / \theta$ than $Q$-dominated models.
Combining such information with Fig. \ref{pinvC7}, we can say that
the cross-spectra of
models with a significant $Q$ component are generally $50 \%$ higher,
or more, than those of
no-$Q$-models, at $k_\parallel u_\perp \, \approxgt \, \pi$. 
This can be understood using our analytic formula in eq.
(\ref{pcrossanalytic}), assuming $n \sim - 2$ to $-3$. 

Put in another way, given an observed auto-spectrum, the inferred
correlation length in physical distance scale is larger in a universe
dominated by $Q$ compared to a no-$Q$-universe; ignoring
redshift-space distortions, this implies a larger cross-correlation
length, and manifests itself in a stronger
cross-correlation for the $Q$-dominated model.
That is what we see in Fig. \ref{pinvC7}.

A good illustration of redshift-space anisotropy 
can be seen by a comparison of the $\Omega_m = 1$ model and the open
model with $\Omega_m = 0.3$ \& $\Omega_k = 0.7$. According to the
dotted line of Fig. \ref{angQK}, the two have very similar $u_\perp$'s
with the latter's a little smaller. This means, with no redshift
distortions, the latter should have a cross-spectrum close to, but 
above, that of the $\Omega_m = 1$ model. Exactly the opposite
is observed in Fig. \ref{pinvC7}. This is because peculiar motion induces
a stronger anisotropy of the correlation function in the $\Omega_m
=1$ model, in the sense of a greater squashing of the correlation length
along the LOS (see eq.
[\ref{Wlinear}]). The end-result is a stronger cross-correlation of 
the critical-matter-density model over the open model. The effect
is not large. It shifts the cross-spectrum of the $\Omega_m = 1$ model
relative to that of the $\Omega_m = 0.3$ - $\Omega_k = 0.7$ model by about
$10 \%$. A similar conclusion holds for most other models in that 
redshift-space distortions change their fractional differences by
about $10 - 20 
\%$. For reference, the values of $f_\Omega$ (eq. [\ref{betaDelta}])
for the 5 models in Fig. \ref{pinvC7} are
$0.98$, $0.77$, $0.78$, $1.0$ and $0.77$ from top to bottom .


It can also be seen from our hypothetical AP test that besides
constraining the absence or presence of $\Omega_Q$, the predicted
cross-spectra can also be used to discriminate between different
equations of state in cases where $\Omega_Q$ is known. 
Unfortunately, according to Fig. \ref{angQK}, the quantity $u_\perp
/\theta$ takes rather similar values for $w \, \approxlt \, -1/3$, which
are also values that seem to be allowed by current supernova observations
(\citenp{garnavich98}). The test does have good discriminating
power among models with larger $w$'s, however.

The distortion parameter $\beta_f$ depends on, in addition to the various
$\Omega$'s, the temperature-density-relation index $\gamma$
(eq.[\ref{betaDelta}]), which while not known precisely, has been shown 
by Hui \& Gnedin \cite*{hg97} to generally lie in the range 
$1.3 \, \approxlt
\, \gamma \, \approxlt \, 1.6$. Varying $\gamma$ within this range
changes the cross-spectra by $\approxlt \, 10 \%$. 

\section{Discussion}
\label{conclude}

At this point, a natural question to ask is: how well
can we measure the cross-spectrum? And given the expected
signal-to-noise of such a measurement, what is the
expected discriminating power among different cosmologies?

The expected variance in
the measured cross-spectrum for a single $k_\parallel$-mode is 
approximately
given by $({P^f_\parallel})^2 + ({P^f_\times})^2$ (see
\citenp{hui98b}). This is obtained by ignoring shot-noise, which is
likely to be accurate because the Poisson-variance is known to be 
sub-dominant compared to sample-variance (at least for the brighter quasars,
see e.g. \citenp{croft98}), 
and assuming Gaussianity, which is probably a good approximation by
the central-limit-theorem, especially on large scales (\citenp{hui98b}). 
Since $P^f_\parallel \gg P^f_\times$ in general, we can approximate
the signal-to-noise for each wave-mode of the measured cross-spectrum by
$P^f_\times / P^f_\parallel$.

\begin{figure}[htb]
\centerline{\psfig{figure=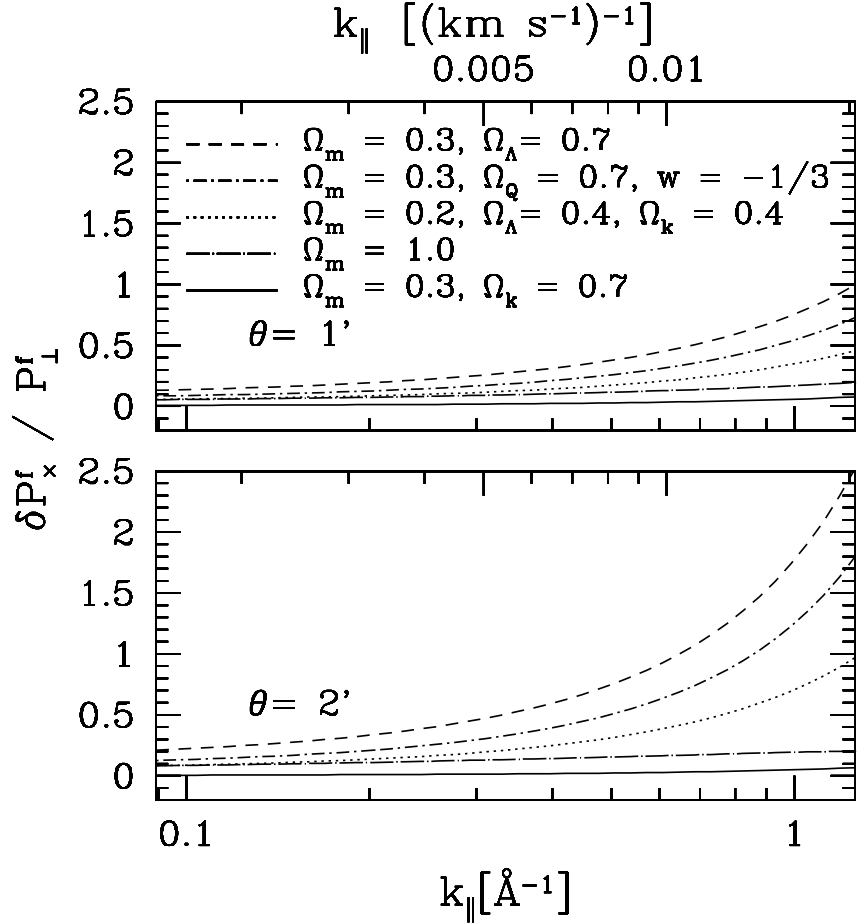,height=5.0in}}
\caption{The fractional error in the predicted cross-spectra (each
panel for the cross-spectrum of the given angular separation) for
5 different assumed cosmologies, labeled according to the order of the curves
from top to bottom. The input cosmology coincides with the assumed
model of the solid line. 
The fractional error is defined as $\delta P^f_\times /
\left.{P^f_\times}\right|_{\rm input}$, where $\delta P^f_\times =
P^f_\times -
\left.{P^f_\times}\right|_{\rm input} $ and $P^f_\times$ is the
predicted/output cross-spectrum. $\gamma$ is assumed to be the
same as the input value (see text).
}
\label{pinvC7}
\end{figure}

On the other hand, our calculation in the last section tells us
what $\delta P^f_\times / P^f_\times$ (Fig. \ref{pinvC7}) is between any
pair of cosmological models. Let us denote these two models by $A$ and
$B$. As argued in \S \ref{AP}, the dominant contribution to
the difference in $P^f_\times$'s can be estimated by ignoring
redshift-space distortions. In that case, the difference in the
cross-spectra is due entirely to the difference in the transverse
velocity-separations (eq. [\ref{uperp}]), let us call them $u_\perp^A$
and $u_\perp^B$ respectively. Assuming $n = -3$, eq.
(\ref{pcrossanalytic}) \& (\ref{projection}) allows us to compute both
$\delta P^f_\times / 
P^f_\times$ and $P^f_\times / P^f_\parallel$, from which we 
can deduce the following: one can rule out model $B$,
if $A$ is the true model, with a $\sigma$-level or signal-to-noise ($S/N$) of 
\begin{equation}
{S\over N} = \sqrt{
\sum_{k_\parallel} \left[{\rm exp}[-k_\parallel(u_\perp^B -
u_\perp^A)]-1\right]^2 {\rm exp}[-2k_\parallel u_\perp^A]}
\label{SN}
\end{equation}
where the sum is over all $k_\parallel$ modes for which the AP test
can be applied. From Fig. \ref{pinvC7}, we will take $0.002 \le
k_\parallel \le 0.02 {\rm \, s/km}$, where the lower limit is set by the
scale at which the slowly-fluctuating
continuum would contaminate the signal, and the upper limit is set by
the scale at which nonlinear distortions would start
to become important (see \citenp{hui98}). 
One can replace the summation by an integration: $\sum_{k_\parallel}
\rightarrow  (L/\pi) \int 
dk_\parallel$ where $L$ is the length of the absorption spectrum
available. Assuming full coverage between Ly$\alpha$ and 
Ly$\beta$, $L \sim 50000 {\rm \, km / s}$. 

Hence, taking model $A$ to be the $\Omega_m = 0.3$ - $\Omega_k = 0.7$
universe and model $B$ to be the $\Omega_m = 0.3$ - $\Omega_\Lambda =
0.7$ universe, we find that, at $\bar z = 3$, the expected $S/N$ is
$0.8$ and $0.6$ 
respectively for angular separations of $1 '$ and $2'$. It turns
out $S/N$ peaks at about $0.5 '$, reaching $1$, but drops off at
smaller $\theta$'s. (The $S/N$ estimate above should be modified
for sufficiently small $\theta$'s because $P^f_\parallel \gg
P^f_\times$ no longer holds. Also, it changes somewhat with $\bar
z$, $n$ and the $k_\parallel$'s we include, but it provides
a good rough estimate for a reasonable range of parameters.) There are two
competing effects: 
a larger $\theta$ gives a larger difference between models (Fig.
\ref{pinvC7}), but also a smaller $P^f_\times$, hence harder to measure.
Note that this is for only one pair of quasar spectra. To reach a $4-\sigma$
level discrimination, something like $25$ pairs, at angular
separations $0.5' - 2'$, would be required.
Note also that in the above estimate, we have ignored the error
in the prediction of the cross-spectrum from the
observed auto-spectrum. This error is likely to be sub-dominant, particularly
because a very large number of LOS is in principle 
available for measuring the auto-spectrum.
There are roughly $10$ pairs of quasars with existing spectra at the above
angular separations, or slightly larger, for $\bar z > 1$ (see
e.g. \citenp{crotts98} \& ref. therein). Upcoming surveys such as
the AAT 2dF and SDSS are expected to increase this number by at least
an order of magnitude. Some of these will be faint quasars for which
shot-noise might be important.

A few issues should be further explored in the application
of the AP test.
First, it is obvious from the above analysis that 
we could boost the $S/N$ for the smaller angular separations
by extending to higher $k_\parallel$'s. This requires, however,
the modeling of nonlinear redshift distortions. The finger-of-God on small
scales would lower our predicted cross-spectrum. The exponential factor
in eq. (\ref{Wperturb}) could be used for modeling this (e.g.
\citenp{pd94}), but it probably depends on more than simply
the $\Omega$'s we are interested in. Moreover, our large-scale linear
distortion kernel (eq. [\ref{Wlinear}]) should be
checked against simulations for accuracy. It is known
in the case of galaxy-surveys, for instance, that
the linear prediction for redshift distortions could be modified even
on very large scales.
Furthermore, the effective bias factor in the distortion kernel
($2-0.7(\gamma-1)$ in $\beta_f$; eq. [\ref{betaDelta}]) might also deviate
from the linear prediction (\citenp{hui98}).

Second, we have focused on a
particular version of the AP test here, in which we construct a
predicted cross-spectrum based on the observed auto-spectrum. In
practice, one should test the whole inversion procedure from the
auto-spectrum to the cross-spectrum with simulated noisy data,
to guard against possible instabilities.
An alternative would be, for instance, to parameterize $\tilde P^f (k)$
(eq. [\ref{projection}] \& [\ref{projectionC}]) by an amplitude and a
slope, and then fit simultaneously these parameters and the various
$\Omega$'s to match the observed auto- and cross-spectra. 
This could lead to smaller error-bars for the measured $\Omega$'s by
restricting the form of $\tilde P^f (k)$.

\begin{figure}[htb]
\centerline{\psfig{figure=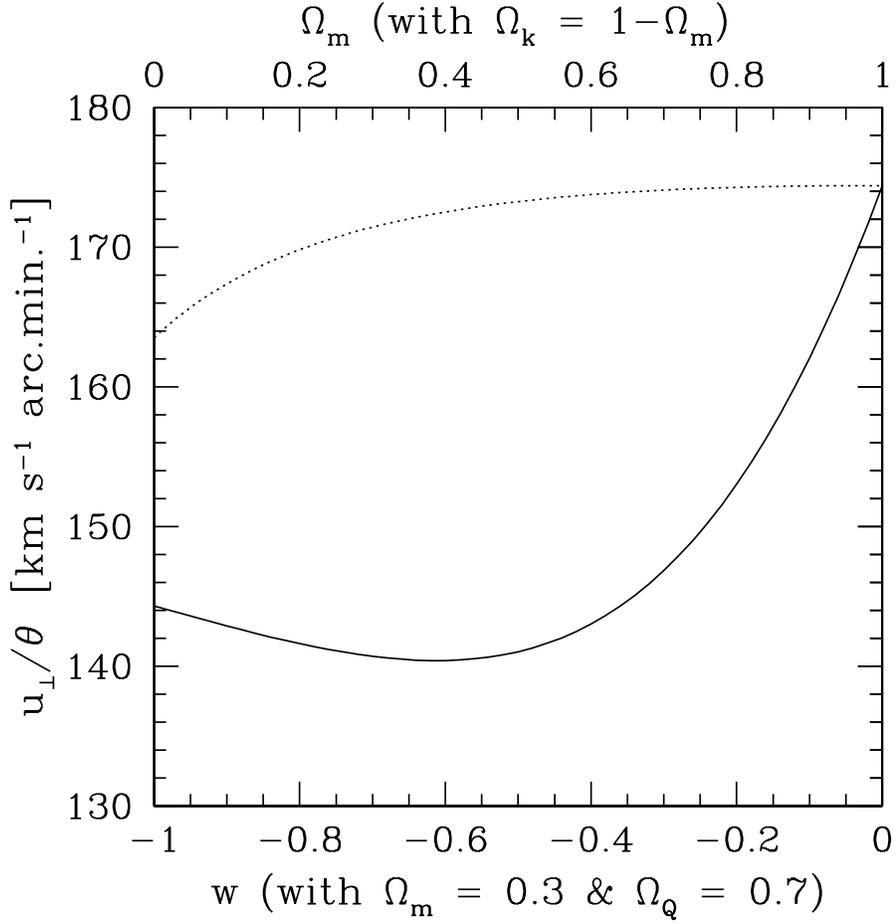,height=5.0in}}
\caption{The velocity-angle ratio (eq. [\ref{uperp}]) at a redshift of
$\bar z = 3$ as a function of two parameters. The bottom solid curve
shows $u_\perp / \theta$ as a function of $w$ (bottom axis) for models in which
$\Omega_m = 0.3$ \& $\Omega_Q = 0.7$. The upper dotted
curve shows $u_\perp / \theta$ as a function of $\Omega_m$ (top axis)
for models 
with no $Q$, i.e. $\Omega_m + \Omega_k = 1$. Models with mixture of
$\Omega_Q$ and $\Omega_k$ smaller than $0.7$ generally have
$u_\perp /\theta$ somewhere between the dotted and solid lines.}
\label{angQK}
\end{figure}

Lastly, we have chosen to focus on a subset of cosmological parameters
which are deemed to be currently popular (Fig. \ref{angQK}). However,
there exist cosmological models which are less conventional, but
nonetheless not necessarily ruled out by current observations, such
as a universe closed by a large $\Omega_Q$ (e.g. \citenp{kam96}). 
Such models could predict a cross-spectrum so strong
that even one pair of quasar spectra would be sufficient
to rule them out, that is, if they do not describe the
actual universe. We will pursue this and other observational issues
in a separate paper.

As this work was nearing completion, we became aware of 
efforts by several groups who considered similar ideas
(\citenp{rupert98,mm98,seljak98}).
This work was supported by the DOE and the NASA
grant NAG 5-7092 at 
Fermilab.


\begin{thebibliography}{}

\bibitem[\protect\astroncite{{Abramowitz} \& {Stegun}}{1964}]{as64}
{Abramowitz}, M. \& {Stegun}, I.~A., 1964,
\newblock {\em Handbook of Mathematical Functions},
\newblock Dover Publications, Inc.

\bibitem[\protect\astroncite{{Alcock} \& {Paczyn\'nski}}{1979}]{ap79}
{Alcock}, C. \& {Paczyn\'nski}, B., 1979,
\newblock {\em \nat} {\bf 281}, 358

\bibitem[\protect\astroncite{{Ballinger} {\rm et~al.\/}}{1996}]{bph96}
{Ballinger}, W.~E., {Peacock}, J.~A., \& {Heavens}, A.~F., 1996,
\newblock {\em \mnras} {\bf 282}, 877+

\bibitem[\protect\astroncite{{Caldwell} {\rm et~al.\/}}{1997}]{caldwell97}
{Caldwell}, R.~R., Dave, R., \& Steinhardt, P.~J., 1997,
\newblock preprint, astroph-9708069


\bibitem[\protect\astroncite{{Croft}}{1998}]{rupert98}
{Croft}, R. A.~C., 1998,
\newblock private communication

\bibitem[\protect\astroncite{{Croft} {\rm et~al.\/}}{1998}]{croft98}
{Croft}, R. A.~C., {Weinberg}, D.~H., {Katz}, N., \& {Hernquist}, L., 1998,
\newblock {\em \apj} {\bf 495}, 44+

\bibitem[\protect\astroncite{{Crotts} \& {Fang}}{1998}]{crotts98}
{Crotts}, A. P. S. \& {Fang}, Y. 1998,
\newblock {\em \apj} {\bf 502}, 16

\bibitem[\protect\astroncite{{De Laix} \& {Starkman}}{1998}]{delaix98}
{De Laix}, A.~A. \& {Starkman}, G., 1998,
\newblock {\em \apj} {\bf 501}, 427+

\bibitem[\protect\astroncite{{Ferreira} \& {Joyce}}{1998}]{ferreira98}
{Ferreira}, P.~G., \& Joyce, M., 1998,
\newblock {\em {\rm Phys. Rev.}}, {\bf D58}, 23503

\bibitem[\protect\astroncite{{Frieman} {\rm et~al.\/}}{1995}]{frieman95}
{Frieman}, J.~A., Hill, C.~T., Stebbins, A., \& Waga, I., 1995,
\newblock {\em {\rm Phys. Rev. Lett.}}, {\bf 75}, 2077

\bibitem[\protect\astroncite{{Garnavich} {\rm et~al.\/}}{1998}]{garnavich98}
{Garnavich}, P. M.~{\rm et al}., 1998,
\newblock preprint, astroph-9806396

\bibitem[\protect\astroncite{{Hui}}{1998a}]{hui98}
{Hui}, L., 1998a,
\newblock preprint, astro-ph 9707068

\bibitem[\protect\astroncite{{Hui}}{1998b}]{hui98b}
{Hui}, L., 1998b,
\newblock in preparation


\bibitem[\protect\astroncite{{Hui} \& {Gnedin}}{1997}]{hg97}
{Hui}, L. \& {Gnedin}, N.~Y., 1997,
\newblock {\em \mnras} {\bf 292}, 27

\bibitem[\protect\astroncite{{Kaiser}}{1987}]{kaiser87}
{Kaiser}, N., 1987,
\newblock {\em \mnras} {\bf 227}, 1

\bibitem[\protect\astroncite{{Kaiser} \& {Peacock}}{1991}]{kp91}
{Kaiser}, N. \& {Peacock}, J.~A., 1991,
\newblock {\em \apj} {\bf 379}, 482

\bibitem[\protect\astroncite{{Kamionkowski} \& Toumbas}{1996}]{kam96}
{Kamionkowski}, M. \& Toumbas, N., 1996,
\newblock {\em {\rm Phys. Rev. Lett.}} {\bf 77}, 587

\bibitem[\protect\astroncite{{Matsubara} \& {Suto}}{1996}]{matsubara96}
{Matsubara}, T. \& {Suto}, Y., 1996,
\newblock {\em \apjl} {\bf 470}, L1

\bibitem[\protect\astroncite{{McDonald} \& {Miralda-Escude}}{1998}]{mm98}
{McDonald}, P. \& {Miralda-Escude}, J., 1998,
\newblock preprint

\bibitem[\protect\astroncite{{Peacock} \& {Dodds}}{1994}]{pd94}
{Peacock}, J.~A. \& {Dodds}, S.~J., 1994,
\newblock {\em \mnras} {\bf 267}, 1020+

\bibitem[\protect\astroncite{{Perlmutter} {\rm et~al.\/}}{1997}]{perlmutter97}
{Perlmutter}, S.~{\rm et al}., 1997,
\newblock {\em {\rm Poster at American Astronomical Society Meeting}} {\bf
  191}, 8504+

\bibitem[\protect\astroncite{{Popowski} {\rm et~al.\/}}{1998}]{popowski98}
{Popowski}, P.~A., {Weinberg}, D.~H., {Ryden}, B.~S., \& {Osmer}, P.~S., 1998,
\newblock {\em \apj} {\bf 498}, 11+

\bibitem[\protect\astroncite{{Riess} {\rm et~al.\/}}{1998}]{riess98}
{Riess}, A. G.~{\rm et al}., 1998,
\newblock preprint, astroph-9805201

\bibitem[\protect\astroncite{{Ryden}}{1995}]{ryden95}
{Ryden}, B.~S., 1995,
\newblock {\em \apj} {\bf 452}, 25+

\bibitem[\protect\astroncite{{Seljak}}{1998}]{seljak98}
{Seljak}, U., 1998,
\newblock private communication

\bibitem[\protect\astroncite{{Turner} \& White}{1997}]{tw97}
{Turner}, M.~S. \& White, M., 1997,
\newblock {\em {\rm Phys. Rev.}} {\bf D56}, 4439

\end{thebibliography}

\newpage


\end{document}